\documentclass{jaa}
\usepackage{graphicx}
\usepackage{siunitx}
\usepackage{xcolor}
\usepackage{hyperref}
\begin{document}\sloppy
\title{Astronomy and Society: The Road Ahead}

%%author names are separated by comma (,)
%%use \and before the last author name
%%use a * along with the number separated by comma
%% for the  author for correspondence
%%\textsuperscript{number} is used for affiliation
%%\affilOne, \affilTwo etc., upto \affilTwentyfive is possible
%%Please note the first letter after \affil is capitalised in the command
%%

\author{A. Sule\textsuperscript{1, 2, *}, Niruj Mohan Ramanujam\textsuperscript{3}, Moupiya Maji\textsuperscript{2, 4}, S. More\textsuperscript{2, 4},  V. Yadav\textsuperscript{5}, Anand Narayanan\textsuperscript{6}, S. Dhurde\textsuperscript{4}, J. Ganguly\textsuperscript{7}, S. Seetha\textsuperscript{8}, A. M. Srivastava\textsuperscript{9}, B. S. Shylaja\textsuperscript{10} and Y. Wadadekar\textsuperscript{11}}
\affilOne{\textsuperscript{1}Homi Bhabha Centre for Science Education, Mumbai, 400088, India.\\}
\affilTwo{\textsuperscript{2}IAU-OAE Center India\\}
\affilThree{\textsuperscript{3}Indian Institute of Astrophysics, Bengaluru, 560034, India.\\}
\affilFour{\textsuperscript{4}Inter University Centre for Astronomy and Astrophysics, Pune, 411007, India.\\}
\affilFive{\textsuperscript{5}Aryabhatta Research Institute of Observational Sciences, Nainital, 263002, India.\\}
\affilSix{\textsuperscript{6}Indian Institute of Space Science and Technology, Thiruvananthapuram, 695547, India.\\}
\affilSeven{\textsuperscript{7}Regional Science Centre and Planetarium, Calicut 673006, India.\\}
\affilEight{\textsuperscript{8}Raman Research Institute, Bengaluru, 560080, India.\\}
\affilNine{\textsuperscript{9}Institute of Physics, Bhubaneswar, 751005, India.\\}
\affilTen{\textsuperscript{10}Jawaharlal Nehru Planetarium, Bengaluru, 560001, India.\\}
\affilEleven{\textsuperscript{11}National Centre for Radio Astrophysics, Pune, 411007, India.\\}
%%escape two column mode for title, affiliation and abstract
%%by giving \twocolumn command as shown

\twocolumn[{

\maketitle

%%include \corres to print the corresponding author Email id
\corres{anikets@hbcse.tifr.res.in}

%%include \msinfo for
%%manuscript information such as
%%received, revised and accepted dates
%%
\msinfo{26 June 2024}{2 January 2025}

%%abstract
\begin{abstract}
Astronomy, of all the sciences, is possibly the one with the most public appeal across all age groups. This is also evidenced by the existence of a large number of planetaria and amateur astronomy societies, which is unique to the field. Astronomy is known as a `gateway science', with an ability to attract students who then proceed to explore their interest in other STEM fields too. Astronomy's link to society is therefore substantive and diverse. In this white paper, six key areas are analysed, namely outreach and communication, astronomy education, history and heritage, astronomy for development, diversity, and hiring practices for outreach personnel. \\

The current status of each of these areas is described, followed by an analysis of what is needed for the future. A set of recommendations for institutions, funding agencies, and individuals are evolved for each specific area. This work charts out the vision for how the astronomy-society connection should take shape in the future, and attempts to provide a road-map for the various stakeholders involved.

\end{abstract}


}]
%%close the twocolumn escape here

%%include \doinum{number}for the DOI number in the header
%%include \volnum{number} for the volume number in the header
%%include \year{yyyy} for  year of publication in the header
%%include \pgrange{num--num} page range of article in the header
%%include \artcitid{num} for the article citation id
%%include \lp to print last page of the article
%%include \setcounter{page}{pagenum} for the exact starting page of the article

\doinum{12.3456/s78910-011-012-3}
\artcitid{\#\#\#\#}
\volnum{000}
\year{0000}
\pgrange{1--}
\setcounter{page}{1}
\lp{1}

\section{Introduction}

Scientific research has led to unimaginable technological benefits for the society. What is less often noticed is that it also affects the way we place ourselves in the universe, the way we view all that is around us. In short, it affects our world view. This knowledge, which scientific research reveals to us, should not be restricted within the scientific community. This knowledge, born of human endeavour, belongs to all of humanity. Contrary to the common questioning for the need for science in a developing country like India, it is precisely in such a society that science is most needed.  Hence a continuous engagement between scientists and society at large is essential. This white paper looks at the various ways of structuring this engagement in the context of astronomy, which also has the natural advantage of having captured the public imagination.
%deeply enslaved by pseudo-scientific ideas, and superstitions - dont say that! start positively.

Interaction between professional scientists and society should be considered as an integral part of scientific education itself. It is often recognised that explaining a scientific concept to a trained scientist is much easier than explaining to someone from the general public. The latter requires the person to have complete clarity of the subject as well as experience in communication.
   
The necessity for interaction between scientists and society also arises from the fact that any research and development activity is increasingly dependent on the support from the society at large. In recent years, the people's representatives have started factoring in society's approval or disapproval of scientific research and scientists while deciding government policy, and it is society's perception of scientists that translates into future generations favourably looking at scientific research as a profession. Thus, interaction with society is an integral part of any scientist's work. On the other hand, the scientific community is a social group and like any other social group, the protocols and practices of interaction within the community are governed by complexities associated with human behaviour and invisible biases. Thus, it is important for scientific community to reflect on the social norms within their own community as well. 

In the case of astronomy, the interaction with society is even more prominent as astronomy often acts as an exciting ``gateway'' science, the discipline which is full of potential for relatable discoveries, leading to more engagement from general audience. Through astronomy, they can then learn the scientific method and apply it in other branches of science or in other walks of life, thus making an overall positive change.

The International Astronomical Union (IAU) recognises the importance of activities relating to astronomers' interaction with society and has a dedicated division to manage those affairs. Their Division C is mandated to look after activities related to "Outreach, Education and Heritage" and is also the division which works with IAU's four offices, namely Office of Astronomy Outreach (OAO), Office of Young Astronomers (OYA), Office of Astronomy for Education (OAE), and Office of Astronomy for Development (OAD). Taking a cue from this model, our discussion will include most of these dimensions as well as other pertinent social issues like inclusion and diversity.

In Indian context, outreach and education efforts by individual astronomers and research institutions have usually happened on their own initiative, without much coordination with the larger professional communities. The first large scale effort by the Astronomical Society of India (ASI) was the formation of the Public Outreach and Education Committee (POEC) in 2014. Since its formation, POEC has coordinated various national outreach campaigns, developed multi-lingual resources, conducted educational workshops, and has been the face of the professional astronomy community on social media. Later ASI also constituted the Working Group on Gender Equity (WGGE), which has been striving towards adaption of more inclusive practices within the astronomy community.

In this white paper, we have tried to address each of the focus areas such as outreach, education, history, development, and inclusion and diversity as a separate sections, which also include action points specific to those sections. We end by reiterating key recommendations.

\section{Outreach and Communication}
In general, the term `astronomy communication' refers to activities through which scientists attempt to communicate their own research work and important new discoveries in their field to the general public. Parallely, 'outreach' includes other standalone interactions with society through lectures, demonstrations, star parties, quizzes, exhibits, videos, social media etc., which are largely carried out by science educators, communicators and amateur astronomers. These efforts help not just in increasing science literacy in society but are also crucial for the development of scientific temper and in dispelling superstitions. When viewed through this lens, one realises that the aim of the outreach is threefold - to inform people about the ``present'' state of knowledge in science in an accessible language, to inculcate scientific temper in our society, and to inspire the younger generation to consider STEM fields as careers in the future.

\subsection{Communication of in-house research}

Communication of in-house research has increased substantially in the last few years due to various factors. Most research institutes now have full-time staff for communication and outreach, which is laudable. Section \ref{instihr} discusses in detail the need for such staff and opportunities such a setup presents. Press releases based on important research or published papers are routinely issued, partly motivated by the requirements of the funding agencies. Another positive development is that a large part of this communication happens through social media and other informal channels.

Moving forward, the demand for science news as well as institutional requirements for press coverage will only increase, and we need to prepare for this as a community. Some of these ways include:

\begin{itemize}
\item There needs to be regular training programmes on astronomy communication for all members of astronomy community including faculty members, post-doctoral fellows as well as graduate students.
\item A standardised module for astronomy communication should be developed with help of existing outreach staff at various institutes and senior science journalists.
\end{itemize}

\subsection{Role of Science Museums, Planetaria and Amateur Astronomers}

Science museums and planetaria across the country are the largest astronomy communicators for the public, with the larger places routinely catering to lakhs of people every year in total. Many amateur astronomers and amateur astronomy clubs regularly conduct star-parties that are more accessible for general audience. Some of these amateurs also regularly cooperate with planetaria on important astronomical events. However, the interaction between the professional research community and these public ambassadors of astronomy has not progressed much beyond individual examples of collaboration. Planetaria and science centres have a clear mandate for outreach and communication, and have trained personnel for doing so. India already has about 67 planetaria and about 10 more are under various stages of installation. There are also about 50 science museums, with small planetaria, developed by National Council of Science Museums (NCSM). Out of these, 11 science museums have their own fixed dome planetaria and most others have mobile planetaria. This is a huge asset for all of us.

Presently, the limited collaboration of professional astronomers limit the scope for updating the knowledge of the planetarium staff. As a result, planetaria are bound to be dependant on commercial vendors for new sky-theatre shows and related content. Most of these commercially available content are prepared outside the country and hence lack Indian context and content. A few planetaria, which are able to make an effort to design their own shows, find the cost of production to be prohibitively high. An active and vibrant collaboration between research institutes and planetaria is necessary to regularly produce new content for planetarium shows, which also highlights Indian research. 

Besides the large sized fixed-dome planetaria, there are also significant number of mobile planetaria across the country, some of which are operated by outreach cells of institutes, some by planetaria and science museums, and some by private players. Their existence is paramount for extending astronomy outreach to remote regions of the country. The projection technology used in these mobile planetaria is also very diverse depending on their origin and their age. Most of the operators of these mobile planetaria are not content designers themselves and they operate them like black boxes. Generating new localised content for this entire ecosystem is challenging, but with help of staff of fixed dome planetaria, it may be possible to design at least short shows in 3-4 dominant formats for the mobile planetaria. Mobile planetaria are also not economically viable for most stakeholders. Lack of low-cost mobile planetaria that are manufactured in the country is a major challenge.

 Hence, the bonds of professional astronomers with planetarium and amateur astronomy community need to be strengthened. Joint development of moving or permanent exhibits, planetarium shows, telescope-making workshops, topical lectures, basic astronomy lecture courses etc., can form a very effective and regular form of public engagement. These can be achieved in the following ways:

\begin{itemize}
\item There should be planned avenues for planetaria and science centre professionals to engage with pan-India professional astronomy community, e.g., involving them in the annual ASI meetings.
\item A structured framework by which institutions can regularly share their recent research with the planetarium community needs to be built.
\item Regular conferences of planetarium and science centre professionals need to be facilitated by appropriate agencies, and a society for planetaria needs to be supported as well.
\item Improving the engagement opportunities between professional and amateur astronomers is crucial. Large gatherings of professional astronomers, such as the annual ASI meetings, or even a dedicated pro-am meetings, can be utilised for this purpose.
\item Research institutes, especially mega-projects with large data-sets, should collaborate with planetaria to produce content for new shows, including involving new media like VR, AR, sonification, haptic models, etc.
\item Currently, manufacturing of mobile planetaria in India is highly underdeveloped, and imported ones are prohibitively expensive. It is imperative that the production of cheaper mobile planetaria inside India is facilitated.
\end{itemize}

\subsection{Public engagement around astronomical events}

Astronomical events like eclipses, transits, the appearance of comets, and planetary conjunctions have always been the primary occasions to capture the public attention. In this regard, the work done in India over the past few decades had been exemplary, and has also served as an example globally. Here, it must be acknowledged that this has been an effort largely led by independent science communication groups, amateur astronomers, and planetaria, and this is to be expected, given their reach. The POEC of ASI, through the live streaming events of eclipses, multilingual information posters, and coordinating events and activities at the national level, has contributed to greatly elevating the scope of these campaigns in the recent past. Many professional astronomers have been involved in popularising these events for many years as well, and there exists a large number of networks and platforms across the country, e.g. All India Peoples' Science Network (AIPSN), that can organise around celestial events very effectively in scale and scope. Another aspect of communication around celestial events is the need to combat pseudoscience around them, and this is yet another field in which there has been enormous effort from across the country. All of this needs to be built upon for the future, for which the following should be considered:

\begin{itemize}
\item Improve the organisation of such events, especially for the research institutions, and their mutual collaborations. E.g., having a common event calendar, having a framework by which planning can be done ahead of time, and having event-based platforms. 
\item The strength of research institutes in this field lies in the production of reliable quality resource content, using their observatories for augmenting live streams, and the availability of scientists to talk to the public about the event, and all this should be strengthened,
\item Support various online digital tools for these events. E.g, the free android apps for Zero Shadow Day, and solar eclipses, which were developed by the ASI POEC, were extremely helpful to the science communicators and teachers across India and abroad. Even better digital platforms will be needed in the future, and supported financially.
\end{itemize}

\subsection{Media Relations}

While audio-visual science content, especially over social media, is increasingly having a wider reach, traditional print and non-print media still plays a very important role in dissemination of science news. Over the last few years, a welcome change has been that science journalists across the country are in regular contact with many astronomers and science communicators. Astronomers are now regularly quoted in media articles, and appear on TV channels as well, improving the quality of shared information. Communication teams of institutions are also well connected with media personnel. This is further helped by active support from the media cells of funding agencies. However, some of the areas that need attention are mentioned below:
\begin{itemize}
\item Facilitate a one-stop media portal for astronomy news for the country.
\item Explore ways in which private TV channels can be made interested in carrying astronomy and science content (which is currently close to zero). This will also help counter pseudoscience on these portals.
\item Support regular interactions with science journalists as well as astronomy bloggers, podcasters etc (e.g., invite them to astronomy events and provide media kits).
\item Develop astronomy modules for science journalism courses in journalism degrees (e.g. this is currently being done at Asian College of Journalism, Chennai).
\end{itemize}

\subsection{Outreach Tools and Materials}

India is a global leader in low-cost science outreach tools, and this is true of astronomy as well. Many science communication organisations, especially those working in under developed regions in the country have produced a range of astronomy tools as well as multi-lingual textual content. Planetaria, science centres, and outreach teams in institutions have done so as well. However, there is no common repository of these material that can be shared by everyone, and some of them have been irretrievably lost. There has been a substantial increase in the quantity of online content on astronomy over the last few years, and this needs to be supported as well as monitored. There has also been little effort in developing high-tech tools, for economic reasons. However, moving forward, there is a need to enlarge and publicise low-cost tools as well as develop high-end ones for science centres etc. Some steps include

\begin{itemize}
\item Fledgling efforts to start a portal for astronomy outreach content have failed, and it is crucial to implement such a platform. This needs to be a separate project with personnel and a budget and will probably take 2-3 years to fully implement and operate. A suitable open-source license should be used for these materials so that they may be shared and reused without restriction.
\item Many individuals working in small towns across India have been making good quality tools and material by themselves, and they need to be supported and encouraged in an effective way.
\item Some mechanism for ensuring the scientific veracity of these material is needed.
\item Support R\&D into hi-tech astronomy tools and kits, in collaboration with science centres and other organisations (e.g., AR, VR etc).
\end{itemize}

\subsubsection{Outreach using TV and Internet}

Although in-person public outreach events are extremely valuable, by their very nature their reach is limited. In a populous country like India, the best way to maximize their reach is through high-quality astronomy programs broadcast through YouTube, TV, and Radio. Live commentaries of celestial events on the radio continue to dominate, especially in rural areas, where TV and internet have limited scope.

Surveys have shown that for young people, the internet, specifically YouTube, is the major source of learning astronomy content (Maji {\em et al.} 2024). Although there are now many documentaries and YouTube videos available, their quality is inconsistent and in many cases the accuracy of information is debatable. Furthermore, many of these videos are available only in English, which remains inaccessible to a large part of India. In this light, there needs to be a concerted effort to produce a series of high-quality videos for the public that will also be translated into regional languages. 

Traditionally, content produced by science institutions lack mass appeal and we must recognise the need to employ the services of professionals to augment these contents.

The format of digital consumption has changed dramatically over the past few years, as shorter videos, such as reels and shorts, have become the format of choice for the younger audience. In the audio space, podcasts and audio stories have greatly risen in popularity. The astronomy community should bring their outreach efforts to these formats as well. 

This content then needs to be distributed through the internet, science communication initiatives by the Government like India Science TV, and mass media (TV, Radio) channels. Private TV and radio channels carry almost no science content currently, this needs to be remedied through government policy.

\subsubsection{Outreach in Indian Languages}

India is a country with a rich diversity of languages. Astronomy and mathematics go hand in hand and educational material in these subjects can be traced back to several centuries, unlike other branches of science. During colonial times, broader access to science by the larger public was made feasible, possibly for the first time, by the pioneering attempts of a few people in languages spoken and used by the vast peoples of India. These include Raghunatha Chary (1868, 71) in multiple South Indian languages, B. G. Tilak in Marathi (1898), and K. Mukherjee in Bangla (1905) among others. They attempted to introduce basic concepts of mathematical astronomy, in a way that is aligned with modern science. This mode greatly helped the dissemination of knowledge. Many universities arranged special lectures and published small booklets on recent advances in science and the tradition of communicating science in a language people understand has continued ever since. 

Despite the widespread use of English in recent years, the majority of the population are more comfortable accessing information in their own languages. Efforts should be made by the scientific community to present their work in Indian languages through print, audio/visual, and other digital media. The diversity in Indian languages can be addressed through a network of translators, science communicators, and amateur astronomer groups. National bodies like the Positional Astronomy Centre, several research institutes and POEC have shouldered the responsibility of rendering the content in regional languages. In last couple of years, spectacular advances in machine translation using large language models have raised the possibility of AI aiding the human translation effort, however at this time the technology is still in too nascent stage to translate materials in Indian languages in an error-free manner.

At the regional level, we see many agencies involved in conducting workshops for science communicators and publish resources in regional languages. Planetaria and science museums also have exhibits and shows in the local languages. However, there is a need for greater and more sustained support from the professional astronomy community.

%{\color{olive} (1) Give references to published work that show that learning basic science in ones own language produces better outcomes. we need to start with this!  

%(2) instis and poec have now started making multi-lingiual material and shud be encouraged. Govt sc comm agencies  and PSA office now have mamny projects for sc comm in reg langs and more are coming. we shud get on to those.

%(3) Train astro communicators and researchers to communicate astronomy in their mother tongue - this is important. this means speak, and write. (examples of such workshops exist). 

%(4) as a policy, insist that all astro material for public should be multilingual when feasible and give awards for it also. this includes subtitles as well.}

 A number of popular lectures related to astronomy and science in general, are available online, in regional languages. A collection of these, for different languages, should be made available at some appropriate website.
 
%A two-week \textit{Raj Bhasha} program is held every year across the country emphasizing the importance of Hindi. This can be extended to science learning and communication, keeping in view the importance of  scientific temper, as stated in the constitution. Science communication in Indian languages other than Hindi should also be duly recognised and rewarded, in a way similar to the \textit{Raj Bhasha} celebrations. This will further help create a pool for science communicators in different regional languages.
Building on extensive programs by agencies similar to Vigyan Prasar and others to promote science communication in Indian languages, dedicated programs to train people in science writing in various languages, contests, awards, and recognition for such activities, and efforts towards dissemination of such material via schools and libraries need to be taken up by the central and state governmental agencies.

\subsection{Capacity Building for Science Communicators}
%Effective science communication can inspire curiosity, fostering understanding, and catalyzing scientific progress. With a steadfast commitment to advancing the frontiers of astronomy and promoting scientific literacy, we envision a future where we could have a group of people who are competent and trained to engage and proliferate the understanding of science in general and astronomy in particular.
%
% The aim is to equip and build a multilingual community of dedicated people with the skills, knowledge, and resources necessary to communicate complex scientific concepts in an accessible and engaging manner and amplify the reach and impact of astronomical research and education.  These communicators include people of diverse range of skills and experience as has been proven by some enthusiasts who have been doing great work for decades, inspiring younger generation to pick up and proceed.
%
%That said, it becomes imperative that science communication be acknowledged as a skill that needs to be learnt and practised. Such recognition will ensure quality in future science communication.

 Effective science communication for the future requires the development of a much larger multilingual community of dedicated people with the skills, knowledge, and resources necessary to do so. It needs to be recognised that science communication is a specialised skill that needs to be learnt and practised. It also needs to be adequately recognised institutionally.

Research and educational institutions need to provide training in science communication for their students and faculty as an integral part of their development. Institutes can introduce appropriate post-graduate courses in science communication as part of NEP reforms as well. 
%
%Academic institutions can organise trainings through workshops, seminars, and online resources designed to enhance the science communication skills. Topics may include effective storytelling with anecdotes and historical references, visual communication techniques, and strategies for engaging with different audience demographics.
%
%Further, the existing community of communicators will be benefit significantly through community engagement may be enhanced by providing opportunities for networking, mentorship, and knowledge sharing. Facilitating connections between researchers, educators, and communicators can catalyze innovative approaches to science communication and amplify the collective impact.

\subsection{Citizen Science Programmes}

The wide availability of internet connectivity in the country offers the opportunity to offer interesting and scalable citizen science programs to the public. Early efforts in this direction have already commenced, e.g. RAD@Home or the One Million Galaxies project or IASC. Earlier experiences with projects like Galaxy Zoo indicate that such efforts rapidly become very popular and produce high quality data that can be used for scientific analysis by professional astronomers. 

Citizen science projects will become increasingly important as data begin to flow from large projects with Indian involvement such as SKA, LIGO and TMT. All of these mega-projects should devote resources to developing and running appropriate citizen science projects, using the data they will obtain with their facilities. While we wait for these facilities to become operational, citizen science projects using data from precursor telescopes can be initiated immediately. 

An area where we can make a substantial difference is in creating citizen science projects that go beyond just pattern recognition by the participant, and provide avenues for further learning and independent thought. There are very few such citizen science projects in the world.
The campaigns on asteroid hunt and naming exoplanets would be a good starting point to attract the young minds.

\subsection{Multi-Institutional / National Outreach Campaigns}

Some of the successful campaigns of the last decade have shown that a much larger impact can be achieved if a well coordinated campaign is run nation-wide simultaneously. At the same time, it must be noted that the success of such campaigns is finely balanced on the just the right degree of coordination. If the coordination is too lax, the campaign doesn't take root nationally. But if the coordination is too tight and controlling, participating groups feel alienated. Keeping this in mind, ways for national outreach campaigns should be developed.  This could be possibly handled by the POEC or a new entity. It should network with the many independent contributors to astronomy outreach, as well as coordinate and support large-scale projects and activities for everyone to be a part of. Such organised, regular large-scale activities for outreach stakeholders to join in and celebrate the success together have proven valuable in recent years and their far-reaching impact has to be repeated regularly. Regular networking among them can be encouraged and support/opportunities for the same can be provided. The increasing number of astronomy NGOs and Astro-Ed companies can also be included in this context. 

\subsection{Science literacy and scientific temper}

The concept of `scientific temper', may be described as `the search for truth and new knowledge, the refusal to accept anything without testing and trial, the capacity to change previous conclusions in the face of new evidence, the reliance on observed fact and not on pre-conceived theory, the hard discipline of the mind—all this is necessary, not merely for the application of science but for life itself and the solution of its many problems' (Nehru 1946). Promotion of scientific temper has been included in our constitution under article 51A(h) as one of the fundamental duties (See Mahanti 2013 for a review) as well.

Astronomy is in an unique position to advance scientific temper. An important aspect of this relates to beliefs among the public regarding astrology, harmful effects of eclipses, etc. Astronomy practitioners, professionals,  those engaged with outreach, and amateurs alike, have had to deal with this difficult issue in any public interaction, and have done very good work in this regard. However, navigating this terrain, where pseudoscience is called out appropriately while acknowledging that astronomy is also a cultural practice, is not easy and needs careful effort. Highlighting the genuine history of Indian astronomy, including `ethno-astronomy' is also important in this context.

Astronomy is accessible to everyone. It may be the only science that reaches remote rural communities at a very low cost. Therefore, it also serves the dual role of mustering inquisitiveness and encouraging scientific temper. Celestial events like eclipses provide an excellent opportunity to reach out to even remote areas easily. We have come a long way from the curfew-like situation of the solar eclipse of 1980, to guided tours to totality locations in the recent past. However, a section of the media continues to encash on these  superstitions. The panel discussions on the private TV channels are monopolised by astrologers who use astronomical images from observatories to glorify their arguments. The solar eclipse announcements generally go with the warning from ophthalmologists
which are also routinely misused, e.g.,  that dangerous UV rays are emitted
during eclipses. Superstitions at various levels are still persistent – there have been news reports of human sacrifices happening in some pockets during eclipses as recent as 2018 and 2019 - and this needs to be fought against aggressively. Whenever an eclipse occurs, the results of previous such studies should be brought to light as a preventive measure.

 Governments should be lobbied with to actively promote the safe public viewing of eclipses by the entire population. The explanation of eclipses with diagrams is provided in text books at the elementary level and students need to be encouraged to do hands-on activities to internalise these concepts, including role-playing games, and regular observational records of the phases of the Moon as community activities. Teacher training about such misconceptions and superstitions is crucial, and our collective experience in dealing with such issues during past eclipses needs to be collated and used in future trainings.

\subsection{International Platforms: OAO and CAP}

Astronomy is a universal science and many of the problems encountered in the outreach effort are common across countries. Recognizing this, the IAU has formed the \textit{Office of Astronomy Outreach} (OAO), with its headquarters in Tokyo, Japan. The IAU also publishes the \textit{Communicating Astronomy with the Public} (CAP) - a journal that shares best practises in astronomy outreach efforts. A periodic conference on this theme is also organised.

Indian participation in CAP meetings is almost non-existent and the number of Indian articles submitted to CAP journal are also minuscule. A possible way to address this issue, and to build national capacity, is to hold national level CAP meetings at regular intervals. Astronomy communicators can gather and learn to present their work in a formal manner. The most promising projects presented at the national meet should be encouraged to be sent to the international CAP meeting/journal.

An Indian journal featuring articles on astronomy communication practices in India would be very useful as well, which also focuses on efforts undertaken in Indian languages.

\subsection{Action Points}
\begin{itemize}
%\item The practice of scientific research is guided by the scientific method. We need to promote the sociological aspects of science practice to also be guided by scientific method and scientific temper.
\item Going forward, we must debunk myths and misconceptions surrounding astrology and other pseudosciences, providing clear and accessible explanations of the scientific evidence that contradicts these beliefs. Through targeted communication campaigns and fact-checking initiatives, we will challenge the validity of astrological forecasts and promote a more skeptical and discerning approach to media consumption.
\item We need to build collaborations that include educational institutions, planetaria, science centres and science museums, media organizations, and other stakeholders to expand the reach of our science communication efforts. By working collaboratively with diverse partners, we will leverage complementary expertise and resources to maximize the effectiveness of our outreach initiatives.
\item We need to develop science communication skills in larger scientific and education oriented workforce through specialised training programmes, materials and portals. Science communication experts should be upskilled to present their work in rigourous and formal way. Opportunities should be created at national level for presenting science communication work in multiple Indian languages.
\item As more young Indians connect to the internet, the opportunities for innovative citizen science projects will only keep increasing. Such projects, especially those reaching out in Indian languages to their target audience should be encouraged.
\item We need to adapt to newer ways of public engangements and collaborate with professional media creators, journalists, bloggers, pod-casters, and social media influencers to bring out science communication materials, which appeal to the current audience.
\item We need to rigorously evaluate the impact of our science communication activities, seeking feedback from participants and stakeholders to inform continuous improvement. By assessing our effectiveness and refining our strategies based on evidence-based insights, we will ensure that our efforts are responsive to the evolving needs and interests of our target audiences.
\end{itemize}

\section{Astronomy for Education}
%Astronomy is the oldest of all natural sciences. It has the ability to inspire children and elders alike due to its universal appeal and a visceral curiosity to understand our place in the Universe. Astronomy is a hub for significant new discoveries, and has been effective in advancing our understanding of nature through millennia. Astronomy is also inherently multi-disciplinary due to the diverse nature of phenomena as well as tools required to study them. (may be move this to overall intro)

Subjects such as Astronomy which have a broad appeal and reach are ripe to be used for the education and scientific training of youth. This is an essential step towards having a scientifically literate society. Unfortunately, the current use of astronomy for education in the country is quite sub optimal and has a lot of unharnessed potential that can be tapped into at various levels.

It is important to realize the challenge that India faces and thereby the opportunity that awaits. India is home to more than 1.5 million schools, involving approximately 10 million teachers, who cater to an unfathomably large 0.25 billion students, just at the school level. The challenge of reaching such a large number is daunting yet achievable with concrete steps that can help utilize the promise of astronomy as a subject. In doing so, it is important to realize that the distinction between "education" and "outreach" is subtle, but significant. Unlike outreach, which is designed to reach the masses and often involves short term engagements, the primary nature of educational activities necessitates a sustained and long term engagement. Such long term engagements require a constant pursuit towards the goal of developing a deeper understanding of science and to transform conceptual landscapes in the learner's mind. Astronomy can be used to develop essential problem solving skills as well as a broader outlook towards life.

Finally, in addition to the creation of a scientifically literate society, India also needs to harness its ultimate capital, that of skilled human resources. Astronomy, for generations to come, will be developed with international cooperation. The success of India's involvement in international mega-science projects that will come to existence in the coming decade crucially depend upon a new generation of scientists and engineers that will require training in scientific thinking, rigor, and the use of scientific methods. In the following subsections we outline the current status of astronomy education in India and the goals that we need to aim for in the coming decade.

\subsection{Enhancing science education in schools}

At the school level, the current science curriculum from the National Council of Educational Research and Training (NCERT) has a single chapter on astronomy, which covers a wide range of topics, all at once\footnote{NCERT initiated a school syllabus and textbook review and rewriting exercise in late 2023. However, till the time of submission of this white paper, the new versions of the textbooks have not been placed in public domain.}. These are supplemented by short units related to the night sky in the middle school Geography curriculum. However, this is clearly not as comprehensive and sustained an activity as it deserves to be. The central and various state boards of education have been repeatedly requested to include and broaden the astronomy content in the school syllabus, unfortunately, to a very limited success.

\subsubsection{Curriculum development}

In early grades, the role of astronomy in our daily lives can be easily highlighted. There are deep connections between the phases of the moon and various festivals that are celebrated in India. The seasons have profound implications for farming practices. The timings of tides and their relation to the position of the moon are connected with occupations such as seafaring and fishing. The tides are also relevant for the generation of electricity by harnessing their energy. These topics need to be weaved in at various levels to make the role of astronomy clearer in our daily lives. Many of these phenomena also require keen observational skills, meticulous tabulation of these observations, and subsequent skills in inference, which form the basis of the scientific method.

In higher grades, as more advanced concepts in astronomy and physics are introduced, their connections with cutting-edge exciting science need to be made clearer. For example, lessons on optics can be connected to the working of space telescopes such as the Hubble and the James Webb Space Telescopes. Images from these facilities already have a wide reach and recognition. Lessons on interference can include mention of gravitational wave detectors and of LIGO-India. Many such advances are expected to happen contemporaneously which require a rapid action task force that can reach a large network of teachers. Learning by students, facilitated by knowledge empowered teachers, can result in stronger and long-lasting education that can keep students motivated and interested in science.

Astronomy in education can not only introduce the field to school students but also serve as a gateway to many other STEM areas. It is clear that the astronomy content at the school level thus needs a revamp and rethinking which will make learning much more meaningful. The results of astronomy research in the Indian context is essential towards achieving this goal (see Sec.~\ref{sec:aer}).

With inputs from research into astronomy education, a model curriculum and lesson plans for school going children need to be designed at the national level. In the process of curriculum design, particular emphasis should be directed towards integrating cultural and local knowledge while concurrently addressing prevalent challenges concerning misinformation, misconceptions, and pseudo-scientific beliefs.
%While designing the curriculum, specific attention can be given towards cultural and local knowledge, at the same time actively addressing issues related to misinformation, misconceptions as well as pseudo-science.

\subsubsection{Resources for Practical Astronomy}
The importance of doing practical work in science education cannot be overstated. This is especially true for astronomy education. Historically, astronomy has been primarily an observational science and many astronomy concepts require an abstract visualization of complex systems. Practical work in astronomy consists of a few different components: direct sky observations and using astronomy simulators.

{\it Use of telescopes:} An early introduction to sky observations using telescopes has immense value to engage students, as it helps in opening 
young minds to the wonders of the universe. Also, such hands-on experience can break the barrier typically termed as 'alienation from experiments', where instruments are viewed as black boxes. However, in India telescopes are available only in a handful of educational institutes, and very rarely in schools and colleges. In a recent nationwide survey, which had half its sample drawn from urban schools, only about a quarter of school students reported having ever observed through a telescope (Maji {\em et al}, 2024). Even when they are available, they are rarely used for sky viewing sessions owing to a lack of training and personnel. This situation needs to be remedied soon. 

Every educational institution should aim to have at least one moderate size (4-6 inch aperture) manual telescope, so that students can learn to enjoy the night sky as well as learn to operate the telescope, from assembly to locating and tracking celestial objects. Building on this experience, institutions with the financial capacity, or a cluster of government funded institutions in a region, should aim to acquire a moderately sophisticated and computerised telescope for more serious amateur astronomy work, which allows them to undertake small science projects as well. Along with this, students should be taught about the night sky, constellations, different types of stars, etc., including the Indian names for them. There should be trained personnel available to operate and maintain these telescopes and train the students and teachers in the schools, as this is a major barrier to telescope access today. Investments in having such positions and in procuring telescopes will certainly prove their worth in time.

{\it Use of computers:} 
There are many dedicated online simulators and software platforms available these days that facilitate scientifically accurate depictions of astronomical systems which can be extremely helpful to visualize and comprehend the key concepts in Astronomy. Stellarium, PhET etc. are some such tools that can be used to enhance classroom learning significantly. Although the computing resources in many schools are scarce at present, the new education policy (NEP 2020) aims to ensure adequate computer resources to each school by 2025. Access to such software needs to be further supplemented by dedicated lesson plans and workbooks for each of the topics that teachers can readily use. There needs to be dedicated teacher training sessions for using these tools and impressing upon them their importance of such tools.

\subsection{Nurturing and encouraging talent}

In science education, one size seldom fits all. Every student has a different comprehension as well as a learning pace. Differential learning approaches have been developed by many researchers, tailored to the needs of the students, which recognize this diversity. Adopting such approaches for astronomy education is a challenge. Thus, in regular classrooms, curiosity of the scholastically advanced students cannot be satiated and need for separate talent nurture programme arises. In this regard, Astronomy Olympiads have been playing an important role in identifying and nurturing talent for Astronomy at the middle and high school level.

\subsubsection{Astronomy Olympiad}

The International Olympiad movement is aimed at bringing the most gifted secondary and higher secondary students of the world together in a friendly competition of the highest level. The Olympiad competitions, including one in astronomy (International Olympiad on Astronomy and Astrophysics - IOAA), provide a stimulus to begin a career in basic sciences or mathematics, and are the meeting places of the brightest young minds of the world. Many friendships forged at the Olympiads form the seeds of scientific collaborations later in life. The Homi Bhabha Centre for Science Education (HBCSE-TIFR) is the nodal centre for this programme within India.

The Astronomy Olympiad starts with a screening Level 1 test where the mathematics and physics skills of a large pool of students are put to test. A select $500$ students are invited to participate in the Indian National Astronomy Olympiad, an exam designed to assess conceptual understanding, logical reasoning, laboratory skills, and above all, the ability to apply problem-solving skills to novel situations, both theoretical and experimental. A 3 week long program trains the top 50 students for various theoretical as well as experimental skills for solving astrophysics problems. The 5 finalists then participate in the International Olympiad with their international peers. Every year the astronomy Olympiad entrants from India win medals in the international competitions. Some of these students take up careers as researchers in basic sciences and also act as mentors for future students. The track record of career choices of Astronomy Olympiad medalists of past 25 years show that more than 50\% of them pursue a Ph.D. in sciences or engineering.

Although the Olympiad screens students and finds the best amongst them, the overall preparation carried out by students nationwide even for the Level 1 examinations often acts as a rising tide that affects many students. The question banks developed by the Olympiads, as well as the resource and training material provided to the top 50 students should be made available publicly, in order to help school teachers and students nationwide.

Given the usefulness of Olympiads, there is a need to generate awareness of its existence amongst school teachers. Conscious efforts need to be put to increase its visibility and to encourage participation of students from semi-urban and rural areas as well as to increase the participation and success rates in a manner that promotes diversity and inclusion along axes such as gender and socioeconomic backgrounds.

There are also other Olympiads which connect directly or indirectly to Astronomy. Participation and selection of students in the IOAA-Jr. is coordinated by the National Council of Science Museums (NCSM) while that in the International Earth Science Olympiad (IESO) is coordinated by Geological Society of India.

Finally, although many of the students selected for the training camps may not take up career in basic sciences, they gain an excellent understanding of the science and the scientific method. Efforts to train and motivate these students to conduct science outreach activities could pay rich dividends to the society. Thus, irrespective of the field they choose in the future, some of these students may end up carrying out science outreach activities in their spare time.

\subsubsection{Astro-internships for skill development}

Almost every research institution in India offers internships and summer programs for graduate and undergraduate students, but none for school students. Some planetaria and science centres do offer such opportunities. Introducing school students to astronomy data oriented projects can aid in deepening their interest and understanding of astronomy, along with developing their computing and critical thinking skills. National Education Policy 2020 guidelines emphasize the need to expose students to computing experiences from an early age. It also urges for a transition from content-based learning to activity-based and collaborative learning. Keeping both of these in mind, an array of astronomy data analysis projects could be introduced to students via an astro-internship program. 

Astronomy research generates a truly huge amount of data regularly, and a wide range of such data is currently available in open-access domains, complete with lesson plans, interactive displays, and online tutorials. Since the last few years, India has been a partner in a number of mega-science projects, and each of them will make their data public along with analysis tools and tutorials. Although some Indian observatories and research institutes make their data available to the public, it needs to be made accessible and usable to interested students. Such data-intensive project internships mentored by astronomy researchers can help develop the computing skills of students which will be extremely useful in their future careers, regardless of whether they pursue astronomy.

\subsection{Research in Astronomy Education} \label{sec:aer}

The development of astronomy curricula in schools must be informed by rigorous astronomy education research (AER). Such studies are routinely conducted in several countries around the world. However, the field is still in a nascent stage in India. Only a couple of Ph.D. theses in the area of astronomy education in India exist and the number of active researchers in this field is quite small, especially given the enormity of the task at hand. These numbers have to increase many fold in order to cater to the large diversity of the Indian school environments. 

To propose changes to the school curriculum, a group of dedicated education researchers and astronomers 
%that looks into the school curriculum throughout the country and identify the places where astronomy could be potentially introduced or integrated into the existing curriculum, 
is essential. Such a group of researchers should systematically define pre-requisites for each concept and create concept maps to illustrate learning trajectories. Such tools can make learning much more meaningful. 
%could be tasked to first identify the topics in astronomy that need to be covered at the school level. These topics can then be represented using concept maps which are diagrammatic representations of the concepts involved to understand that given topic. Given the expected level of mathematical and cognitive skills of students, such concept maps are a powerful way to plan and design an age-appropriate curriculum. The existence of these concept maps could also serve as a road map for teachers in order to plan their instruction. For students, such concept maps help them to gauge their understanding of the subject as well as to make connections to the different concepts they know or have learnt about. 

Many AER studies over the years (e.g. Lelliott \& Rollnick 2010, Vosniadou \& Brewer 1992, 1994) have shown that there are numerous misconceptions about a variety of astronomical phenomena that are closely held by students (e.g. the misconception that seasons occur due to a change in the Sun-Earth distance throughout the year) and if these alternative ideas are not directly mentioned and confronted while teaching, even after learning the correct theory, students will eventually revert back to their misconception. Recently, a baseline survey of astronomy education was conducted in India with high school students (class IX), which revealed that although students are generally highly interested in learning astronomy, there are notable deficiencies in students’ understanding of basic astronomical concepts, with only a minority demonstrating proficiency in key areas such as celestial sizes, distances, and lunar phases (Maji {\em et al.} 2024).

The results from AER studies can have a significant positive impact on the teaching methods of astronomy concepts. Unfortunately, there is often a vast disconnect between research and its implementation. This picture needs to be changed at the earliest.
%To remedy this, the results from AER should be distilled down to a summary that is understandable by non-specialists, and it should be made  easily accessible to the general public. Furthermore, lessons from AER should be translated into a list of actionable points and such ideas need to reach teachers directly.

These activities require a closer connection between institutes concerning science education and those engaged in astronomy research. Such initiatives have just started to emerge -- the International Astronomical Union's Office of Astronomy for Education (OAE) Center India was established in the year 2022, as a collaboration between the Inter-University Center for Astronomy and Astrophysics (IUCAA) and the Homi Bhabha Center for science education (HBCSE). Similarly, the Indian Institute of Astrophysics (IIA) has started its Science Communication, Public Outreach and Education (SCOPE) initiative as well as the Cosmology Education and Research Training Centre (COSMOS-Mysuru) in collaboration with the University of Mysuru. There needs to be increased coordination between such initiatives, such that regional research in Astronomy education can be spread throughout the country.

\subsection{Aiding teachers}

\subsubsection{Teacher training programs}

Various surveys have found that school teachers often have to teach subjects that are not their area of specialization. Consequently, astronomy concepts are often taught by teachers who do not have any background in astronomy or physical sciences, and struggle to teach the astronomy content. Hence, effective teacher training programmes are urgently required. 

Currently, there are several teacher training programs conducted by various research institutions around the country, but the number is minuscule compared with the need for it. Additionally, there is no coordination among these programs and there is no quality control to maintain a set standard. There should be a set of guidelines drawn up for such training workshops along with a repository of handbooks, lesson plans, worksheets, activity guides, and multimedia resources that can be shared with the teachers. The training materials in the repository should be translated into various Indian regional languages as a large fraction of teachers are mainly comfortable in their own language and this is often a barrier to accessing quality teaching resources.

Such training programs should also include ideas on integrating innovative teaching practices in classrooms while teaching astronomy, such as the use of technology tools, online resources, project-based learning, and game-based learning to make the learning experience innovative and fun. Crucially, the number of training programs should be scaled up significantly and there should be adequate funds and staff allocated for conducting these programs regularly. 

\subsubsection{Pedagogy repository}

While in-person teacher training sessions can significantly help teachers, it may not always be accessible for all teachers countrywide. Thus, the in-person training programs should be complemented by an online central repository website containing the best teaching methods for different astronomical concepts that can be accessed by any teacher. Such a repository should maintain updated information about the teaching methods, detail any classroom activity that can enhance the lesson, and list other additional resources e.g. images, videos, simulators, etc that can aid the teacher. The repository should also contain creative situation-based questions, known as concept inventories, that can better probe the student's understanding of the concepts compared to the traditional fact-based questions. 
%Teachers can use ideas from these inventories to test their students at the end of lessons. This repository should also have a feedback process where teachers can report back what other methods they have tried in their class, what works and what does not in a regional context. 
The content on this website needs to be multilingual and well organized with appropriate tags. The teachers' handbooks should be accompanied with appropriate links to such content.
%, and there should be a provision to include some of the newer developments in astronomy. The creation of such an interactive website can become an extremely helpful tool to improve pedagogical practices. 

\subsubsection{Assessment tools}

To gauge the success of any educational initiative, such as teacher training or curricular materials, it is important to perform a timely assessment. Assessment for such programs are of three types, front-end evaluation (done before the program, often to know the expectation of participants),  formative evaluation (done during the program), and summative evaluation (done at the end). These can test for various things, e.g. if their awareness or knowledge increased, if their interest grew or attitude changed, if they gained skill, etc. There should be such evaluations done after each initiative through either a set of questionnaires, interviews, or a mix of them. The curricular materials can also be tested in the same way by comparing the pre and post-test results using statistical methods. In some cases, we may be interested to know if the participants retain the knowledge long-term, and for these, a follow-up assessment needs to be made. Guidelines for how to conduct these assessments should be prepared, each program should be assessed, and changes should be made periodically based on the results of assessments.

\subsection{Action items}

\begin{itemize}
    \item Despite its considerable potential, the integration of Astronomy into school science lessons remains underdeveloped. Drawing from research in astronomy education, a national-level model curriculum and lesson plans should be formulated to capitalize on the inherent interest students have in astronomy and its utility in teaching other scientific disciplines.

    \item Schools should be equipped with telescopes and provided with trained personnel or appropriate training for their effective use, alongside the integration of online tools to enhance astronomy education.

    \item For nurturing talents across the country, we need to increase awareness and participation in specialized programs like Olympiads among school teachers and students, especially in rural areas, while encouraging science outreach activities for students selected for training camps of the same. 
    
    \item Astro-internship programs that can provide motivated students with hands-on experience in various astronomy data analysis projects, which will enhance students' computing skills, irrespective of their future career paths need to be implemented.
    
    \item Astronomy education research (AER) in India is its infancy, necessitating heightened research endeavors, particularly through increased funding and the establishment of research programs. In the short term, we also need to form a specialized group comprising education researchers and astronomers to identify avenues for integrating astronomy into the school curriculum.

    \item There is a compelling need to significantly expand the scope of teacher training programs with a focus on innovative teaching methods and the provision of multilingual educational resources accessible via online platforms. Additionally, implementation of timely assessments to gauge the effectiveness of each educational initiative is essential to ascertain the efficacy of educational interventions and facilitate informed adjustments based on evaluation outcomes.
\end{itemize}

\section{History and Heritage}
The curiosity in the human mind about celestial bodies is evidenced by the countless stories and myths in cultures throughout the world. India has a long tradition of careful observations and modelling of the night sky. Our society is replete with examples of cultural practices, scholarly works and monuments which have firm astronomical roots. There are multiple isolated examples of historians of science and astronomers looking at different aspects of this vast canvas, out of their own interest. However, no university or research institute or a department in them, is dedicated to the study. Therefore the work done so far seems to lack coherence. We first present a list of different aspects which may warrant careful research before making our recommendations:

\subsection{Material Heritage}
Material heritage in astronomy comprises tangible objects (e.g. astrolabes), structures (e.g. monuments) and documents (e.g. manuscripts) which can be analysed with near-unanimous conclusions. This is true with all cultures and many astronomical concepts hitherto unknown are being brought to light (Hoffmann \& Wolfschmidt 2022). The 400 year old astrolabes from India lie scattered all over the globe in museums and private collections. Fortunately, a comprehensive catalogue is now available (Sarma, 2018) after decades of work. Similarly, other simpler gadgets like water clocks, sun dials at Harappan sites and elsewhere, and the megalithic stone alignments, need to be studied with great care and patience. 

\subsubsection{Astronomy in Manuscripts and Documents}
Many people associate Indian astronomical tradition only with manuscripts in Sanskrit. There is a long list of dedicated astronomical treatises in Sanskrit which start from {\it Vedang Jyotish} of {\it Lagadha}, written probably 3500 years ago till the texts of Pathani Samanta Chandrasekhar in the 19\textsuperscript{th} century. The obvious highlight of this tradition is the {\it Siddhantic} Astronomy era which lasted for more than 1000 years between 5\textsuperscript{th} and 15\textsuperscript{th} centuries had a number of luminaries such as {\it Aryabhata, Varahamihira, Brahmagupta, Lalla, Bhaskara, Madhava, Nilakantha etc} contributing novel and unique techniques to accurately compute observed coordinates of the sun, moon and the planets. Outside these dedicated texts, we also see a number of astronomical references concealed in other texts such as {\it Puranas, Ramayana} and {\it Mahabharata} and many literary works in Sanskrit as well as other regional languages. The first source book appeared in 1985 (Subbarayappa \& Sarma, 1985) and subsequently very few comprehensive books on history of Indian astronomy (Ramasubramanian {\em et al.} 2016; Ram \& Ramakalyani 2022) have been published.

It is on a positive note, that we see a number of scholars rigorously engaged  in  the last century towards a faithful translation of the texts to contemporary languages and decode the numbers and equations hidden in verses. This approach also uses astronomical references to date the texts themselves. They have also provided a mechanism to  explain comprehensive models of the solar system as envisioned by the original authors. This area of research, however, is in need of a meta-analysis of all previous works to compare them with each other for consistency and to bring out a coherent picture. 

At the same time, one should note that there are astronomical texts written in other Indian languages dating from a few centuries ago to probably more than a thousand years. These texts are based on the same sanskrit astronomy texts, translated with or without commentaries in other languages. They are yet to be studied and hence the numbers do not compare with Sanskrit texts. Thus, there is a clear need to urgently collect, document and preserve these texts from different parts of India and study them for their astronomy content. There have been some sporadic efforts to study texts in Malayalam, Kannada, as well as Persian language texts from Mughal courts, but they are far from comprehensive. Attempts to do so for Kannada have been started recently by the IIA COSMOS project in Mysuru. Similarly, star catalogues, though only a handful in number, need to be catalogued and published. (Ram \& Ramakalyani 2022). 

The astronomical history extends to the colonial period also. Fortunately several records are available in legible formats; they need to be documented before those references are lost. Be it Jesuit priests of Goa in the 17\textsuperscript{th} century or Le Gentil in the 18\textsuperscript{th} century or Pogson and others in the 19\textsuperscript{th} century, all form an important part of this country’s astronomical history. Most of the colonial documentation, on a parallel study of traditional Indian astronomy is already lost. Similarly, the 20\textsuperscript{th} century history of astronomical institution building and telescope making should be preserved for posterity.

An often neglected aspect of astronomy knowledge from our region is that of navigation by the fisher and trading communities (Arunachalam 2002). These communities produced complex navigation tables and methods, along with novel instruments to be used at sea. The methods used also differ widely as we go along the Indian coast (e.g. regions closer to the equator which cannot see the pole star have an entirely different system of navigation). This needs further study and documentation, since this knowledge will be lost very soon.

\subsubsection{Astronomy in Structures and Buildings}
Observatories of Sawai Jai Singh II, popularly known as Jantar Mantar, are India’s most significant astronomical monuments. The observatory in Jaipur is in excellent condition, and the ones in New Delhi and Ujjain are mostly in acceptable condition. But the one in Varanasi need more attention to restore it to its original state. Studies highlighting the importance of these instruments, (Sharma 2016) and efforts to create awareness (e.g. the campaign by N. Rathnasree with school children) have been partly successful. 

There are many temples and caves scattered across India, which have incorporated astronomical elements in their design and construction (Shylaja \& Geetha 2016). The most important task ahead is to delink astronomical significance from myths and exaggerated legends. A proper study and documentation of these monuments is necessary as these buildings in public domain are potential sites to engage society with astronomy.

Apart from these, there are other sites associated with medieval period and colonial eras which have astronomical connections. Some examples are Pari Mahal in Srinagar or Pogson’s grave in Chennai or sites of solar eclipse observations in Guntur (1868) and in Vijayadurga (1898). The details on some defunct observatories and planetariums like the one in Mujaffarpur and Pune also need to be documented. All these sites should be developed for astronomy outreach connecting the site’s history to present day astronomy. Even the Central Station of India at \ang{82.5}E and \ang{23;11;}N (for fixing the IST) needs to be developed as an educational centre for schools and the general public.

\subsubsection{Astronomy in Artefacts}
In the Mughal and Bahamani periods, a number of astrolabes were created in India with dials marked in Persian and/or Sanskrit. These astrolabes lying in different museums in India and abroad are testament to the level of refinement of astronomical observation skills of our ancestors. For other types of small instruments, available in museums, documentation need to be prepared on their use and importance. 

Records of kings’ donations found on copper plates and stone inscriptions usually cite the astronomical event associated with their dates. These are useful records of the events as well as the contemporary astronomical tradition and beliefs of the society (Shylaja \& Geetha 2016). This information may not always be in words. Sometimes it is in the form of drawings etched or painted on stones or painted as murals on walls of buildings.

Similarly, old instruments of Madras Observatory (now IIA), UP state observatory (now ARIES) etc. are important markers of our astronomical history which provide us avenues of engagement with wider society. Forgotten places like the planetarium and observatory in Muzaffarpur also need to be revived. 

\subsection{Cultural Heritage}

\subsubsection{Astronomy in calendars, {\it panchanga} and physical calendrical markers}

Indians follow a variety of calendars. Lunisolar calendars like {\it Vikram Samvat} and {\it Saka Samvat} are popular in some parts of the country. Many calendar makers follow {\it Suryasiddhanta} but its principles are modified to suit the local needs.  Many of these calendars also have their own {\it panchangas} (almanac of planetary and asterism positions). In addition, many indigenous communities have their own calendars. Many of them also use local geographical markers to keep track of sunrise / sunset positions and hence overall season cycles. These markers may be a geographical feature such as a hill or an old tree or sometimes they can be in the form of specially positioned megaliths along required line of sight. In the Indian context, the documentation of such marker alignments as well as comparative study of {\it panchangas} or calendars has been done by a couple of researchers over the past decade but that has barely scratched the surface. Just south of {\it Vindhyas}, there are 200+ megalithic sites and only a handful of them have been explored for their possible astronomical alignments (Menon 2019).

\subsubsection{Astronomy in Star Lores and Myths}
Although many indigenous communities have not preserved their astronomical culture in written documents, they are traceable in their oral history. The trade routes along the seacoast also demanded a good knowledge of the night sky. They have their own stories for shapes of constellations and interesting names for other astronomical phenomena like eclipses, lunar halo and the like. Over the last decade researchers have collected such star lores from about 10 out of more than 700 indigenous communities around the country (Shetye {\em et al.} 2023). These data collection attempts reveal an alarming situation that the oral history of these communities is being lost with the passing away of the older generation. Thus, an urgent pan-Indian effort is needed to document this cultural capital.

\subsection{Recommendations}
\begin{itemize}
\item There is a need to bring coordination and coherence among the handful of researchers in the history of astronomy. A possible way to achieve this will be creation of a ‘virtual hub’ for research in the history of astronomy in the country. The hub should facilitate exchange of ideas and expertise among researchers and organise periodic review meetings for knowledge exchange, for contemplating future directions of research collectively and for kick-starting collaborations.
 \item There are many scholars outside India (for example, Europe, Japan and the US) engaged in the study of astronomy manuscripts and recreating the calculations. The institute hosting this virtual hub should invite selected experts from abroad once in a year or so for discussions and mutual exchange of ideas.
\item Research institutes and funding agencies should announce opportunities for competitive research grants in interdisciplinary areas such as history of astronomy and encourage those interested to explore it systematically.
\item Many manuscripts are available in museums and archival collections like the Oriental Research Institute in Mysuru, Bhandarkar Oriental Research Institute in Pune, L. D. Institute in Ahmedabad and so on. However, the researchers find it extremely difficult to access the records. It would be ideal to have the digitised versions made available.
\item ASI meetings should have a session on the History of astronomy, and separate workshops on the subject need to be supported.
\end{itemize}

\section{Astronomy for Development}
Historically, the night sky has always been an awe-inspiring sight. The increasing light pollution due to the expansion of cities and towns has relegated the access to dark skies to rural areas, mainly the sparsely populated ones. Consequently, a very tiny fraction of the city dwelling population has experienced the night sky in all its glory. On the other hand, the rural population, while accustomed to the night sky, is largely unaware of the scientific know-how related to the night sky and the livelihood opportunities it can provide. Therefore, it is essential to protect the visibility of night sky and educate people about it. 

At the same time, Astronomy practice of research, education and outreach should contribute to sustainable development of the society in social and economic aspects. Astronomy can lead to entrepreneurial activities, astro-tourism and developing new pathways in other knowledge domains.

The International Astronomical Union (IAU) and the South African National Research Foundation (NRF) have jointly established the Office of Astronomy for Development (OAD) in 2011 with the mission to use Astronomy as a tool for societal development by funding various projects. While setting up such a dedicated body may not be feasible for a single country like India, there is a lot of scope for action.

\subsection{Development Inclusive of Light Protected Zones}

\subsubsection{Sky Darkness at Observatory sites}
 At present, most of the optical astronomical observatories in India possess acceptable dark skies for observations (limiting magnitude of 5.0 or better). However, that may not be the case in future with increasing population, access to electricity and light in rural areas and tourism in hilly regions as observed around some of the older observatories, e.g. Japal Rangapur Observatory (JRO). Particulate matter content in atmosphere over Indian subcontient is typically much higher than cities in developed countries, leading to worse astronomical seeing. 

 Such pollutants cannot be controlled by mere government regulation. Confidence building measures with local population and financial assistance to mitigate financial costs associated with changed practices has yielded better results in similar situations in other countries. Similar efforts are needed to maintain sky darkness in the regions around existing major Indian optical facilities at Hanle, Devasthal, Kavalur, Udaipur, etc. Measures like providing correctly shaped night shades to cover street lighting are commonly pursued by observatories because a direct causal link is apparent. But efforts such as educational activities with local schools leading to a population sensitive to the need of better night sky or helping set up new technology alternatives to discourage crop burning or providing particulate matter filters to local businesses are indirect measures which work towards long term sustainability of the observatory site.

 Radio observatories encounter similar problems due to radio interference from human activities. Radio interference can be cause by high tension power lines, electric railway routes, mobile phone transmissions, industrial machinery, etc. Efforts of different kinds such as providing financial assistance for installation of Faraday cages as necessary would be needed to protect areas around GMRT, Ooty and Gauribidanur. 

\subsubsection{Dark Sky Sanctuaries}
Around the world, there are several designated 'dark sky sanctuaries' (DSS), where special measures are taken to keep light pollution to a minimum. Establishment of India’s first dark sky region, the Hanle Dark Sky Reserve (HDSR) centred around the Indian Astronomical Observatory (IAO) and covering an area of over 1000 square km in Ladakh, is a fantastic first step in the right direction. More sites should be identified around the country which can be bestowed with DSS status and with the help of respective local governments they can be developed as regions with pristine dark skies.

\subsubsection{Dark Sky for Urban Outreach}
Increasing light pollution is also a hindrance to science outreach activities in urban areas. The limiting sky magnitude in core urban areas is just 2.0 or even worse. Due to density of urban population and extreme vehicular pollution, any improvement in those conditions is extremely challenging. However, it may still possible to identify some areas in the vicinity of urban settlements, which still have reasonably dark skies (limiting magnitude of 4.0 or better). Efforts should be made to develop these areas along the principles of DSS guidelines, albeit with slightly relaxed norms. If such custom developed dark sky parks are within a short driving distance of metro cities, they would become hot spots for star parties leading to more local employment opportunities.
 
\subsection{Astronomy for Livelihood}
\subsubsection{Astro-Tourism} 
India boasts of a small but sizable amateur astronomy community. Some of these amateur astronomers have been conducting stargazing sessions for the public using telescopes for many years. Such activities along with astrophotography images inundating social media have created interest in tourism with Astronomy as one of the focal activities (Astro-Tourism for short). Those opting for astro-tourism travel to remote locations to experience clear dark skies. A quick online search will reveal quite a few advertisements about stargazing events in various parts of India. However, most of these events are held within a couple of hours of driving distance from major cities, where sky quality is only marginally better than city skies. Availability of skilled human resources with suitable telescopes and safety concerns at remote locations still pose a challenge and have kept the potential of Astro-Tourism restricted so far. If skilled personnel from rural areas are trained and sites suitable for Astro-Tourism are developed, then it can provide in-situ job opportunities and aid in reducing migration for jobs, particularly from remote hilly areas. 

There are a few individuals who work as freelancing sky experts for luxury resorts in locations such as Lakshdweep and Andamans. Given the large tourism industry in India and neighbouring countries, there is a lot of potential for growth in this sector. 

\subsubsection{Astro Start-ups}
Opening up the space sector to private players is going to increase commercial activities in space. This is already amply demonstrated in USA with companies like SpaceX and now new age Indian start-ups are also entering this niche sector (e.g. the recently successful rocket launch by Agnikul Cosmos). Many entrepreneurs are using this opportunity to create start-ups in the space sector and allied areas. Engaging with such start-ups would be a win-win situation as it will not only contribute to their growth, but also provide employment opportunities for those young astronomers who are interested in switching over from academia to industry.

Most of the good quality telescopes acquired by institutes, universities, science centres, planetaria, amateur astronomy groups and individual astronomy enthusiasts for educational and astronomy outreach purposes are imported from out of the country. There have been small scale entrepreneurs who have tried to manufacture/assemble small telescopes locally in the past, but were unable to sustain it for long. Additionally, these telescopes were lagging in quality as compared to the imported ones. In recent years, a few manufacturers have started efforts to make telescopes in India. Such manufacturers would require support and encouragement to excel and match the finesse of international brands.

There are a few start-ups in metro cities in ed-tech sector specialising in astronomy education. These companies conduct lessons and activities in different schools and also mentor citizen science activities with student participants. Similarly, small private players in mobile planetarium space are a crucial last mile resource to take astronomy to those remote schools where big institutions fail to reach. The recently organised conference by the Indian Institute of Astrophysics on "Astro-tourism and astro-entrepreneurship in India" was an attempt to bring this fledgling community together.

\subsubsection{Transfering Astronomy Skills to other Domains}
The skills learnt by Astronomy graduates, especially those related to big data and machine learning, are transferable to many other disciplines from economic policy to bioinformatics. However, this aspect of transferability of skills is never emphasised in traditional educational system, which is where a change in outlook is needed. The Pune Knowledge Cluster has started many projects recently, some of which use the tools and methodologies in astronomy to solve problems in the health and life sciences sectors.
%
%A few institutes and universities have initiated astrobiology programmes, which are in nascent stages. Engagement by professional astronomers will be helpful in the growth of such programmes and development of astrobiology research in the country. Awareness of the tools/ experiments/ trials of experiments in space conditions from the domain of life sciences, psychology, physical nature can be incorporated and embedded as examples and illustrations in the existing texts, manuals etc so that learners as well as others associated with such subjects get better insights in such studies and the knowledge evolves as interdisciplinary.

\subsection{Action Points}
\begin{itemize}
%\item Standardised short training modules in telescope handling and night sky watching should be developed and there should be coordinated pan-India effort to teach these skills to youth in areas with dark skies.
%\item Technical support in the form of distribution of cost-effective telescopes or telescope making workshops, where the telescopes made may be provided to the participants, should be made available to the populations living in DSS or DSP. There should be also long term support available for refurbishing/cleaning/coating of their telescopes.
\item Dark sky sanctuaries (DSS), along the lines of the Hanle Dark Sky Reserve, should be created around every major optical/IR observatory and Dark sky parks (DSP) should be created close to major urban centres and tourism hotspots for astronomy outreach. These should be designed to provide livelihood opportunities for local communities both as providers of science knowledge and logistics services.
\item There should be efforts to increase awareness among the hotel/tourism industry in remote areas about Astro-Tourism and its potential as a revenue source. This includes standardised short training modules in telescope handling and night sky watching for local populations in DSS or DSP. 
%After this training, astro tourism initiative can be supported not just by distributing telescopes but also post-distribution technical support for troubleshooting / cleaning / coating / refurbishing of their telescopes.
\item Projects like DSS or DSP cannot succeed without active support from local administration. Thus, there should be regular special drives to increase awareness among local administration about Astro-Tourism and technical support in developing suitable sites (like was done in Ladakh). 
\item Association of Indian universities and institutes or department of higher education can be approached to incorporate a section of a course on ‘Astro-tourism’ as a part of their courses on tourism studies, so that learners who take up such courses on tourism and aspire to become tourism professionals can become aware of the idea. The HDSR has started doing so already.
\item Steps should be taken to provide technical guidance and encouragement to Indian telescope makers and other start-ups, to improve quality of their products and make them competitive in the domestic and world market. Manufacturing in India offers a significant price advantage and if quality is matched, these businesses can easily capture a lion's share in the market.
\item Awareness about light pollution and radio interference should be created through outreach programmes. Further involving the industry (the design section) in those campaigns so that when they develop and design such outdoor fittings, they keep the light pollution aspect in mind. This can also serve as an academia-industry collaboration which will in turn benefit the society.
%\item Light pollution maps of regions around various optical observatories should be showcased and highlighted at every available opportunity. Night Time Light Atlas by National Remote Sensing Centre (NRSC) may be used for this purpose.
\item Entrepreneurship in developing small dome planetariums at the block level countrywide to be encouraged to proliferate modern astronomical thoughts.
\item Astronomy can impart skills which can be utilised effectively in other knowledge domains for overall economic and social development of society. Such transferrability should be highlighted in the training of astronomy students.
\end{itemize}
%\textcolor{red}{Again, this AfD section is too focussed only on astro tourism and dark sky, and that is very limiting. One subsection to add is to use astro for getting underprivileged children back to school and to like learning, etc etc.}

\section{Diversity, Equity, Inclusivity}
There are two broad aspects to diversity, equity and inclusion (DEI) in the context of astronomy. First, the practice of astronomy, like any other science, is a social activity and is affected by the inequalities and exclusions that society at large suffers from. As a community driven by the scientific method and motivated by scientific temper, we need to develop an inclusive, diverse, and equitable culture within scientific institutions, associations, and communities (see e.g. chapter 7 of STIP). Second, the subject matter of astronomy, with its very scale and grandeur, is uniquely suited for promoting concepts of inclusion, empathy, and humanity among the public at large. There is tremendous scope for utilising this aspect of astronomy in the country.

The National Education Policy 2020, the National Curriculum Framework 2005 before it, and other documents talk of equality of access, opportunity, as well as outcomes (see e.g. sections 6.1 and 14 of NEP 2020 or the NCF 2005). These need to be reflected within our academic and research environments, and in the context of this white paper, within astronomy communities and practice. This would include all outreach and education events, sky watching activities, planetaria and science centres, workshops, public talks, and so on. The Working Group on Gender Equity (WGGE), set up by the ASI, has been a torchbearer in Indian science in the area of gender. Learning from the experience of WGGE, the ASI needs to expand its focus on DEI to other under-represented communities as well, in the coming decade. 

\subsection{Building a representative and diverse astronomy community}

For many decades, attempts at promoting diversity of the scientific (and corporate) workforce had been countered with the bogey of fall in merit. However, a number of studies worldwide (Adams 2013, Freeman and Huang 2014, Nielson {\em et al.} 2017; AlShebli {\em et al.} 2018) have shown that increasing diversity and equity leads to an increase in the quality of the work produced by the community. Concomitantly, it is also our constitutional duty to ensure the inclusion of traditionally and historically under-represented groups of people and thereby include the skills they bring with them (e.g., Preamble, Indian Constitution). We aspire to build an Indian astronomy community that is welcoming to diversity and enables a friendly and supporting work environment for everybody. There is a large body of evidence that the practice of science worldwide is still affected by unconscious biases (and sometimes conscious bias and harassment as well). The antidote to unconscious bias, as discussed by social scientists, is to acknowledge that scientists are not immune to it, and then to put in place specific practices to reduce it (e.g. Aloisi and Reid 2020 in the context of astronomy). A well known example is of dual-anonymised refereeing by the HST Time Allocation Committee, which was shown (Johnson \& Kirk 2020) to substantially increase the fraction of accepted proposals from women astronomers (and now being adopted STScI for JWST, ESO, SARAO, etc). Similar conclusions have been arrived at by double blind refereeing on journal articles in terms of gender (Caplar et al 2017, Ni et al 2021), race and country of origin (Yousefi-Nooraie et al 2006) of authors, etc. We have no reason to believe that the Indian community would not be subject to similar trends in unconscious bias.

Apart from gender, which WGGE has been leading the efforts in, the Indian astronomy community has not yet started the conversation on the lack of diversity in terms of caste, region, religion, gender identities and special abilities, which are some of the major markers of exclusion in the country. Some of these discussions will be difficult and may even be viewed as polarising. However, there are many governmental and non-governmental groups who have decades of experience in navigating these issues, who can help us in this regard. 

Lastly, the concept of the 'astronomy community' also needs to be diversified. Astronomy instrument builders, engineers, astronomy communicators, telescope operators, computing experts, planetarium staff, etc, all of whom make the practice of astronomy possible, should be seen as a part of the ASI community, and be included in meaningful ways.

\subsection{Quantifying diversity and inclusion audits}
The first step in this direction is to collect and publish quality statistics and estimate the level of diversity. There have been many studies with regard to gender (see \ref{links} for links to previous reports). This needs to be done for multiple identities that can marginalize people, viz. caste, region, religion, sexuality and gender identity, disability, urban vs rural, nature of higher education background, etc. Some of these axes apart from gender were indeed surveyed in a few instances but not in a substantial way. WGGE has been collecting some of these statistics meticulously over the years, which can serve as a model.
% (see https://astron-soc.in/wgge/sites/default/files/pdfs/Statistics_2021.pdf)

Most of our institutions were built decades ago, including many of our observatories. Although providing access is now mandatory for all new constructions, very few existing facilities are disability friendly. A common practice to ensure access is to invite advocacy groups to conduct disability audits, which should be encouraged.

An obstacle faced by minority group members in any environment is that of bullying, harassment, etc, and only some of this can be amenable to administrative or legal remedy. Apart from this, many communities face specific structural issues (e.g. transgender scientists changing their names on publications and linking to past ones with their dead names, gender non-conforming scientists and students using safe or preferred washrooms, seminar halls with no lift access, resting place for people with invisible disabilities, etc)\footnote{We can take inspiration from the American Astronomical Society which has set up separate committees to look into issues of gender, race, disability, sexuality and gender identity, etc}.

\subsection{Inclusive access to knowledge} 

Inclusive access to science is an active field of work globally, and this focuses on how scientific knowledge can be accessed by people who have visual, auditory, and other challenges. In the last few years, a few groups in India have been doing work in this area as well. 

There is now a global community of astronomy educators who are working on tactile or haptic models in astronomy for students with visual impairment and other disabilities, including sonification projects (Casado {\em et al.} 2021, Arnio {\em et al.} 2019, Ortiz-Gil {\em et al.} 2011a, Ortiz-Gil {\em et al.} 2011b). India has a few manufacturers of haptic material in education, which we need to tap into. In addition, routine practices like subtitles and alt-text on all video material, transcription of lectures, and guidelines for making slides and plots colour-blindness friendly, are easy to implement and should be made standard practice in the community. To achieve this, astronomers and astronomy educators should be sensitised towards need of such practices.

In the area of public outreach, the last few years has seen a large growth in production of resource material in multiple Indian languages, which is to be lauded, and strongly encouraged in the coming years. Statistics also show that the participants in ASI meetings are becoming more diverse in terms of location of work, university participation etc (see Secretary's Report, ASI 2022), and this needs to be monitored and supported. For example, the recent practice of including a weekend in the ASI meeting dates has benefited many University staff in being able to attend.

\subsection{Designing inclusive astronomy events and conferences}

To make any event an inclusive one requires planning and forethought, and best practices need to consciously incorporated into the function of the LOC, SOC, etc for all astronomy events. An inclusive event is one where (1) there are minimal structural barriers for someone to apply to and attend the event and (2) where all attendees are able to participate in all of the planned activities there. 

For example, physical locations for the event need to be accessible to people with disabilities. Night sky watching activities in outreach events typically see a low attendance from women students due to apprehensions of safety, and involving the families in such activities can help in this regard. Hybrid conferences during the pandemic have clearly shown a substantial increase in the diversity of participants, which can be due to economic or personal constraints on travel. It is tempting to make the organising and planning of outreach events completely online for our ease, but this does not take into account the digital divide in our society (e.g. registration only via google forms leads to less response from rural communities). Academic conferences can be intimidating for newcomers, especially students, and a structured response to this problem would help quite a bit. Similarly, having family friendly events especially with childcare, multi-lingual events when relevant, etc also help inclusion. Bullying and harassment at these events also need a mechanism of reporting and redressal.

From these few examples, it is clear that there are a number of practices that we should adopt as a matter of policy, to enable inclusive and accommodating astronomy spaces. This should take the form of guidelines and codes of conduct that organisers of any event should follow. The ASI could take a lead in formulating such documents for the astronomy community, after examining similar codes of conduct  already put in place by the International Astronomical Union (IAU), Americal Astronomical Society (AAS), Indian Academy of Sciences (IAS), Indian Social Science Council (ISSC), etc.

\subsection{Astronomy, shared humanity, and global citizenship}

Astronomy is enormously inspirational, and hence its popularity amongst people at large compared to any other branch of science. This inspirational value arises from the existential nature of the questions asked by researchers, the sheer scale of size, mass, and time involved, and the unfamiliar or even exotic aspects of celestial objects. Astronomy as a subject illustrates the vastness of the universe, existence of other worlds and even hope for discovery of life elsewhere. Outreach programs routinely benefit from these advantages, and highlight them as well. However, a conscious incorporation of these values in these programs needs the practitioners to appreciate them, and be provided with the tools to talk about them. A well known example of such a project is that of the 'Pale Blue Dot', a flagship project of IAU OAD, and an IAU 100 project. 

"Universe Awareness" (UNAWE; Miley et al 2020) was a torchbearer in this domain, which came up with a structure and the pedagogy to use astronomy for global citizenship among children around the world (there were a few events in India as a part of UNAWE as well). Following this, many projects funded by the IAU Office of Astronomy for Development are also in this vein (Benitez-Herrera and Gonzalez 2020, Fragkoudi 2020). Other examples are projects in the refugee camps in Nigeria and Uganda (see Sec \ref{links}). Science communication groups in many parts of India talk about shared humanity etc in the context of their astronomy outreach, which can be aided by the production of resource material relevant to the Indian conditions. 

\subsection{Gender Equity}

Due to the work of the Working Group for Gender Equity (WGGE)\footnote{https://astron-soc.in/wgge/about-wgge} of ASI and many individual astronomers, the questions around issues of gender balance, gender bias and discrimination, structural and societal impediments towards gender equity in astronomy, etc. are now routinely discussed in the community, including at the ASI meetings. The WGGE has also been systematically collecting data on gender balance in various institutes in the country and has, along with the ASI, also been conducting gender audits of meetings (Kharb 2020). 

The WGGE needs to be strengthened and supported further, but institutions also need to take on the onus of self-monitoring and awareness raising. Some of the science academies in the country have committees on gender issues, and the astronomy community can be in the forefront of a nation-wide platform in STEM. Importantly, the WGGE is an endorser of the {\it Hyderabad Charter for Gender Equity in Physics} (see Sec \ref{links}), which provides specific recommendations for institutions and departments, conferences, national agencies, and teaching. The astronomy community needs to take these on board for implementation, as well as support a robust mechanism for periodic evaluation and course correction. It is also important to emphasize that the task of promoting gender equity needs to be distributed and carried out by organisations and individuals in their own domains as well.

These efforts can be taken further to also recognise individuals with other gender identities and their involvement in creation of new knowledge and inclusion of issues faced by them in our discussion in gender sensitive practices.

\subsection{Action Points}

\begin{itemize}
\item Support regular collection and publication of statistics on diversity of the astronomy workforce in India in terms of gender, caste, region, disability, gender identity and sexual orientation, etc. Support gender audits of astronomy meetings, region audits of all astronomy meeting attendees, disability audits of astronomy facilities including observatories, etc.

\item Form a national task force along the lines of ASI-WGGE that is tasked with working towards increasing diversity and equity in astronomy along multiple axes. Similarly, encourage formation of DEI (diversity, ethnicity and inclusion) committees in each astronomy institute.

\item Formulate a code of conduct for all astronomy meetings and workshops (there are many examples from other astronomy societies and institutions around the world) which can be a guideline for organisers. Encourage conferences to have an ombudsperson for dealing with problems.

\item The LOC of all meetings to have an inclusive access plan (e.g. all talks to be broadcast online, physical access to venues, etc).

\item Increase the number of fellowships, subsidies, and scholarships to increase participation by astronomers and students, who are in greater need of such support, in national and international meetings.

\item All publicly funded projects should have a mandatory component of outreach and/or community development and should have explicit mandate to be inclusive and strive towards diversity. Suitable changes in policies should be discussed with appropriate funding agencies.

\item Practices, such as double blind peer review, that help reduce unconscious bias should be made mandatory in time allocation committees of Indian astronomical facilities and grant proposal committees. 
\end{itemize}

\section{Special purpose centres at research institutes and dedicated human resources} \label{instihr}

Over the past decades, many astronomy institutes had a few members of their faculty or scientific staff who were keenly interested in Education and Public Outreach (EPO). Some of them went on to establish and nurture outreach programmes for their institutes, and some moved to planetaria and other non-governmental spaces to pursue their passion of bringing astronomy to the public. In the 1990s, IUCAA became the first research institute to have a dedicated full time group for EPO. As of 2024, most major astronomy institutes have hired dedicated EPO personnel. There are regular outreach events on national science day and other special days and in some cases, the footfall for these events is in the tens of thousands. These dedicated personnel also liaise with amateur astronomers, planetaria, science centres etc. creating a conducive eco-system around the institute.

Similarly, outreach groups in several IISERs and IITs are mentored by Astronomy faculty members in those institutes. Astronomy outreach is also a critical part of any large astronomy project, especially those with a footprint on Indian soil such as LIGO-India and TMT-India. EPO activities ought to be seen as a part of academic and social work of institutes, which is also mandated through governmental policy regarding "Scientific Social Responsibility". 

\subsection{Hiring and Career Progression}

So far, each institute has adopted different approaches to devise a solution for effective outreach. The permanent EPO staff may get hired at pay level 12 at some institute and at pay level 6 at another. In some other cases, only project positions are offered. Some institutes focus on getting younger personnel at the entry level positions and some facilitate lateral entry of experienced people from non-governmental spaces. Lateral entry of those with outreach / education / journalism experience at appropriate senior level is highly desired. However, present hiring norms for senior level entry do not allow for this possibilty at all. For example, for direct recruitment at senior level in most scientific institutes, a Ph.D. in Science is mandatory and that requirement eliminates most candidates with invaluable field experience.

There is a pertinent need to bring coherence to hiring practices and a balance between youth and experience, to optimise the potential of an outreach team. There should also be clarity on the career progression route of the outreach personnel. The EPO projects handled by these personnel should have their own evaluation parameters, which would naturally differ from the standard parameters used to evaluate astronomical research work. Stagnation in career often leads to stagnation in work.

\subsection{Training and Networking}

For young recruits, including project personnel as part of outreach teams, there are not many possibilities of pre-service training. A good outreach / education person would not just need command over basic astrophysics, but also requires good communication skills in English and the regional language of the state, a good eye for design aesthetics, a sense of pedagogy / teaching skills, and familiarity with the night sky and telescopes. Typically, it is expected that they will also pick up additional skills like graphic designing, video editing etc on the fly.

Having a good astrophysics training in UG / PG or being an amateur astronomer during student days may help these potential recruits in this regard, but more structured training is needed before one can effectively face a crowd and win them over. Thus, there is a clear need for a training process for new recruits. As the number of EPO staff at each individual institute is only a handful, it makes immense sense for the institutes to pool in resources and organise combined training of their outreach teams. The EPO groups in these institutes can be further trained to document history of astronomy in their region and also curate history of the institute in which they are working. Institutes should provide dedicated budget for such programmes.

Networking of all outreach teams across institutes to work on pan-India projects and also to share resources and materials is crucial for producing more results with smaller staff strength. Institutes should facilitate visits of their outreach personnel to other institutes for exchange of ideas and collaborative work.

\subsection{Action Points}
\begin{itemize}
\item Institutes which don't have dedicated EPO cells yet, should be encouraged and guided to form the same. Convergent policies should be evolved for hiring and training of EPO staff.
\item Create flexible hiring policies for lateral entries at senior level to attract those with field experience.
\item Develop a customised protocol for assessment of work by the EPO staff at the institutes and ensure smooth career progression.
\item Coordinate dedicated training workshops for amateur as well as professional astronomers, including graduate students and post-doctoral fellows, on how to communicate science effectively.
\item There is currently no formal course on astronomy communication offered anywhere in the country, which is a huge lacuna that should be addressed, either through UG level electives in a few places or as continuing education certificate courses. 
\end{itemize}

\vspace{-2em}
\section{Discussions and Summary}

Humans have been interested in the sky since pre-history, and hence astronomy, as a body of knowledge, belongs to all of humanity. Astronomers and astronomy institutions therefore have an obligation to share their work with the public, in collaboration with a multitude of stakeholders like planetaria, science centres, amateur astronomers, science communicators, science teachers, students, etc. Professional astronomy institutions have a key role to play, especially in consonance with the Astronomical Society of India (ASI). 

In this white paper, an analysis of these areas has resulted in a number of recommendations for each of them, which were discussed in the previous sections. The main stakeholder of this exercise being the ASI, many of these recommendations are for the Society, a large part of which is to develop national level portals to collate, organise, and distribute key information. These include portals for sharing research and new developments with planetaria and science centres, as well as with the media, database for multilingual outreach and communication material from diverse organisations and individuals, a portal to consolidate material on history of Indian astronomy, model syllabi and curricula for all educational levels, and databases for various astronomy service providers. The ASI also needs to lead efforts to enhance astronomy in school and college education, coordinate training on astronomy communication for its members, establish a task force to push for substantive inclusion and diversity in astronomy on multiple axes, and continue support to the ASI Working Group for Gender Equity. 

Institutes, universities, and their staff have the onus of increasing their access to the public in the form of dissemination of their work especially in multiple languages and modern digital formats, open up their campuses in appropriate ways, develop innovative ways of engaging with socio-economically disadvantages communities, evolve assessable policies on diversity, equity, and inclusion internally, hire dedicated staff for communication and outreach with well defined career paths, and evolve avenues for research in history of astronomy.

The astronomy community is larger and more diverse than the membership of the annual meeting attendees of the ASI, and the various practitioners of astronomy needs to find a place in the ASI. The community at large needs to evolve a consensus on open data policies, ethics and codes of conduct, and foreground scientific method and scientific temper in their internal and public engagement.

Lastly, funding agencies need to evolve norms for hiring, supporting and funding outreach practitioners and programs, especially through dedicated grants, develop policies and metrics for assessing institutional and project based outreach efforts, and promote introduction of science communication courses in universities.

As we plan for the future of Indian astronomy, our vision as a community will need to engage with the public in ways that are substantive, inclusive, and diverse. The recommendations in this work hopefully provide a guide to such a path. However, progress depends on proper planning, execution, and periodic re-evaluation, which the ASI is best placed to lead.

%
%
%%Appendix
%
%\appendix
%
%\section{An appendix section}
%Text goes here (Radhakrishnan 1980).
%\begin{equation}
%x=a+b+c
%\end{equation}
%
%\vspace{-2em}
%
%\section{Another appendix section}
%Text goes here.
%\begin{equation}
%y^2=ax+b+c
%\end{equation}
%\vspace{-3em}
%
%%Use section* for acknowledgements
\section*{Acknowledgements}
The authors acknowledge extensive discussions over many years with people science movements and planetarium personnel in shaping many of these ideas.

\vspace{-1em}

%%use \balance somewhere in the left column of the last page to balance the two columns in the end page

%%References section

\section*{Useful Links} \label{links}

\begin{enumerate}
\item STIP, Science, Technology, and Innovation Policy 2020, Ministry of Science and Technology,
{\tiny \url{https://dst.gov.in/sites/default/files/STIP\_Doc\_1.4\_Dec2020.pdf}}
\item NEP 2020, National Education Policy 2020, {\tiny \url{https://www.education.gov.in/sites/upload\_files/mhrd/files/NEP\_Final\_English\_0.pdf}}
\item NCF 2005, National Curriculum Framework 2005,  
{\tiny \url{https://ncert.nic.in/pdf/nc-framework/nf2005-english.pdf}}\\

 \balance
\item Report of National Task Force for Women in Science, DST, {\tiny \url{https://www.ias.ac.in/public/Resources/Initiatives/Women\_in\_Science/taskforce\_report.pdf}}

\item 'Science Career for Indian Women' 2004, INSA Report, {\tiny \url{https://www.ias.ac.in/public/Resources/Initiatives/Women\_in\_Science/report.pdf}}

\item IAS-NIAS Research Report 2010, {\tiny \url{https://www.ias.ac.in/public/Resources/Initiatives/Women\_in\_Science/surveyreport\_web.pdf}}

\item Vision Document from Inter-Acacdemy Panel 2016, {\tiny \url{http://insaindia.res.in/pdf/ws\_anx\_1c.pdf}}

\item Godbole \& Ramaswamy, {\tiny \url{https://www.ias.ac.in/public/Resources/Initiatives/Women\_in\_Science/AASSA\_India.pdf}}

\item NITI Aayog Report, {\tiny \url{https://vdocuments.net/2016-17-mallick-president-dr-rajlakshmi-mallik-project-director-sri-siladitya.html?page=1}}

\item IAU 100: Pale Blue Dot, {\tiny \url{https://www.iau-100.org/pale-blue-dot}}

\item IAU OAD Pale Blue Dot, {\tiny \url{https://astro4dev.eu/projects/pale-blue-dot}}
 
\item UNAWE, {\tiny \url{https://www.unawe.org/resources/reports/}}

\item Nigerian refugee camp project, {\tiny \url{https://astro4equity.org/space-and-stem-camps-for-refugees-in-nigeria/}} 

\item Ugandan refugee camp project, {\tiny \url{https://www.astro4dev.org/knowledge-access-and-sharing-through-cultural-astronomy-in-ugandas-refugee-settlements-and-host-communities/}}

\item Hyderabad Charter for Gender Equity in Physics, {\tiny \url{https://astron-soc.in/wgge/sites/default/files/pdfs/HyderabadCharterForGenderEquityInPhysics.pdf}}

\end{enumerate}

\end{document}